\documentclass[draft]{agujournal2019}
\usepackage{url} 
\usepackage{lineno}
\usepackage[inline]{trackchanges} 
\usepackage{soul}

\newcommand{\cels}{\,^\circ\mathrm{C}}
\newcommand{\psu}{\,\mathrm{g \, kg^{-1}}}
\newcommand{\sfx}{\,\mathrm{g \, m^{-2} \, s^{-1}}}
\DeclareUnicodeCharacter{2217}{$\ast$} 

\usepackage{amsmath}

\usepackage{ragged2e}
\justifying

\draftfalse

\journalname{JGR: Oceans}

\begin{document}

%
%


\title{Bistability of ice-shelf cavity circulation and melt under coastal polynya forcing}

%
%




\authors{Louis Saddier\affil{1}, Corentin Herbert\affil{2}, Christopher Y. S. Bull\affil{3}, and Louis-Alexandre Couston\affil{4}}


\affiliation{1}{ENS de Lyon, CNRS, LPENSL, UMR5672, 69342, Lyon cedex 07, France}
\affiliation{2}{CNRS, ENS de Lyon, LPENSL, UMR5672, 69342, Lyon cedex 07, France}
\affiliation{3}{Australia's Climate Simulator (ACCESS-NRI), Australian National University, Canberra, Australia}
\affiliation{4}{Université Claude Bernard Lyon 1, ENS de Lyon, CNRS, LPENSL, UMR5672, 69342 Lyon cedex 07, France}




\correspondingauthor{Louis Saddier}{louis.saddier@ens-lyon.fr}



\begin{keypoints}
\item A 3D idealized model of ice-shelf cavity captures irreversible polynya-induced transitions between cold and warm circulation regimes
\item Thermocline depth at the ice front is an indicator of changes of melt and circulation regimes
\item Transition irreversibility depends on offshore water properties and modeling choices affecting local dynamics
\end{keypoints}

%
%

%
%


\begin{abstract}
Antarctic ice shelves play a key role in controlling the mass loss of the Antarctic Ice Sheet.
Increasing evidence from regional studies suggests the possibility of abrupt transitions from low to high basal melt regimes under future climate forcing.
The computational cost and complexity of these realistic simulations yet limit their ability to systematically investigate the steady-state response of ice shelves to incremental forcing, the reversibility of abrupt transitions and the underlying positive feedback loops.
Here, we report a detailed analysis of abrupt regime changes in an ice-shelf cavity with idealized geometry, subject to prescribed temperature and salinity stratifications in the far field and near-ice-front surface salt fluxes representing the impact of sea-ice production in coastal polynyas.
We show that a cold regime emerges when the buoyancy loss in the polynya is sufficient to trigger deep convection, while a warm regime develops otherwise.
The steady response of the system exhibits hysteresis, characteristic of irreversible transitions, and scaling laws are proposed for both melt rates and transition thresholds. 
We demonstrate that bistability emerges when a gyre recirculates the meltwater within the ice-shelf front area. The bistable forcing range increases with far-field ocean temperature but decreases with background stratification. We also show that increasing horizontal viscosity limits bistability, which may explain the range of behaviors reported in previous studies.

\end{abstract}

\section*{Plain Language Summary}
Ice shelves are the floating extensions of the grounded Antarctic Ice Sheet. 
Their thinning, which is controlled by their basal melt rates through interactions with the ocean, can unlock rapid grounded-ice discharge into the ocean and thus lead to increased sea-level rise. 
Recent studies suggest that some ice shelves may abruptly shift from low to high melting rates and may not easily return to their initial state.
Here, we use a simplified numerical model to investigate why such transitions occur and what controls their reversibility.
Our model resolves the dynamics of an ice-shelf ocean cavity and includes the effect of sea-ice formation near the Antarctic coast.
Surface waters become dense and mix with deeper waters when sea-ice formation is sufficiently intense, bringing cold water beneath the ice shelf and maintaining low melting. 
However, when sea-ice formation weakens, warmer ocean water can enter beneath the ice shelf, causing higher melting.
We find that both states can exist under the same sea-ice formation rate, meaning that a transition to high melting may not reverse when initial conditions are restored. 
This behavior strongly depends on ocean circulation near the ice shelf and on the temperature and salinity of the offshore ocean.


%
%

%


%
%
%
%

\section{Introduction}

Antarctic ice shelves lose mass primarily through calving and basal melting~\cite{galton-fenzi2025AntarcticIceSheet}.
The recent acceleration of ice-shelf thinning~\cite{paolo2015VolumeLossAntarctic} has increased the discharge of ice and meltwater into the ocean~\cite{furst2016SafetyBandAntarctic}, enlarging the Antarctic Ice Sheet contributions to sea-level rise~\cite{galton-fenzi2025AntarcticIceSheet} and unlocking possible modifications of the global ocean circulation in the near future~\cite{li2023AbyssalOceanOverturning}.
Although most of the observed Antarctic mass loss is currently concentrated in West Antarctica, ice-shelf thinning is highly heterogeneous across the continent~\cite{paolo2015VolumeLossAntarctic,galton-fenzi2025AntarcticIceSheet}. Critically, the comparatively stable East Antarctic ice shelves buttress a larger reservoir of grounded ice, suggesting substantial potential for future mass-loss acceleration if presently stable regions become destabilized.
These regional contrasts are largely controlled by the properties of water masses intruding the continental shelf~\cite{thompson2018AntarcticSlopeCurrent}, and differences in ocean circulation within ice-shelf cavities~\cite{silvano2016OceanIceShelfInteraction}. 
Although some cavities are filled with cold ($\approx -2^\circ$C) Dense Shelf Water (DSW), formed at the surface freezing point, and are associated with low basal melt rates, others are instead filled with relatively warm ($\approx 1^\circ$C) Circumpolar Deep Water (CDW), originating from the Antarctic Circumpolar Current, leading to high basal melt rates~\cite{silvano2016OceanIceShelfInteraction}.

Coastal polynyas, sea-ice-free regions at the front of the ice shelves, are thought to play a key role in controlling the circulation regime beneath Antarctic ice shelves~\cite{silvano2016OceanIceShelfInteraction,moorman2023CoastalPolynyasEnable,oconnor2025EnhancedWestAntarctic,haigh2026WindDrivenCoastalPolynya,saddier2026IrreversibleTransitionsOcean}. When katabatic winds push sea ice away from the coast, opening polynyas, the exposed ocean surface loses significant heat to the cold polar atmosphere~\cite{noel2025AtmosphericResponseAntarctic,golledge2025AntarcticCoastalPolynyas}. As a result, sea-ice formation rates are the highest in polynyas and lead to significant surface buoyancy losses~\cite{thompson2020FrazilIceGrowth,ushio1993LaboratoryStudySupercooling}, which can trigger deep convection events when sufficiently intense~\cite{ohshima2022DominantFrazilIce,kusahara2010ModelingSeaIce}, and thereby support cold circulation regimes.

Evidence of possible cold-to-warm transitions in ice-shelf cavities has been put forth for past climates~\cite{suganuma2025AntarcticIceshelfCollapse} and regional studies suggest that such transitions may occur under most future climate forcing for the Filchner-Ronne~\cite{hellmer2012TwentyfirstcenturyWarmingLarge,hellmer2017FateSouthernWeddell,hazel2020BistabilityFilchnerRonneIce,daae2020NecessaryConditionsWarm,naughten2021TwotimescaleResponseLarge,haid2023DriversRegimeShifts,hoffman2024IceshelfFreshwaterTriggers,song2025RegionalConditionsDetermine} and Amery ice shelves~\cite{jin2025CurrentReversalLeads,haid2023DriversRegimeShifts}, and within most Antarctic ice-shelf cavities under the SSP5-8.5 scenario~\cite{mathiot2023SouthernOceanWarming,mathiot2026TransitionMuchWarmer}. 

The high computational cost and complexity of forcing schemes in realistic configurations make it difficult to systematically investigate the steady response of an ice-shelf cavity to incremental variations of external conditions.
As a result, a quantitative understanding of the physical mechanisms and positive feedback driving these transitions and their thresholds is lacking.
Specifically, only a limited number of studies have examined the reversibility of these regime transitions. Past works exploring the reversibility of ocean circulations have focused on the Amundsen Sea~\cite{caillet2023DriversReversibilityAbrupt}, the Filchner–Ronne ice shelf~\cite{hazel2020BistabilityFilchnerRonneIce,haid2023DriversRegimeShifts,reese2026IceSheetOceanInteractionsReversibilitya,naughten2026LimitingGlobalWarming,hellmer2017FateSouthernWeddell}, the Ross ice shelf~\cite{naughten2026LimitingGlobalWarming}, and pan-Antarctic configurations~\cite{mathiot2026TransitionMuchWarmer}, but with diverging conclusions.
Although some studies found irreversible dynamics~\cite{hellmer2017FateSouthernWeddell,naughten2026LimitingGlobalWarming,hazel2020BistabilityFilchnerRonneIce} or hints of hysteresis behaviors~\cite{haid2023DriversRegimeShifts} for the case of the Filchner-Ronne ice-shelf cavity, others reported reversible transitions~\cite{mathiot2026TransitionMuchWarmer,reese2026IceSheetOceanInteractionsReversibilitya}.
Quantifying the reversibility of cold-to-warm transitions is yet critical for assessing their potential short and long terms impacts, particularly in scenarios where the external forcing may weaken or reverse, or simply exhibits temporal variability.

Here, we adopt a complementary approach to realistic studies by running simulations of a canonical ice-shelf ocean cavity with idealized boundary forcings implemented within the widely-used NEMO ocean code. We focus on regime transitions linked to changes in surface buoyancy fluxes within polynyas at the ice-shelf front. The simulations enable a systematic investigation of the steady response of the ice-shelf cavity circulation to polynya forcing and of the reversibility of the transitions, thus providing  results from a 3D ocean code complementary to our previous analysis based on a low-dimensional box model~\cite{saddier2026IrreversibleTransitionsOcean}.
Our configuration is inspired from existing idealized ice-shelf cavity configurations~\cite{holland2008ResponseIceShelf,holland2017TransientResponseIce,arthun2013EddyDrivenExchangeOpen,hunter2006SpecificationTestModels}, and more specifically is built upon the model intercomparison project ISOMIP+~\cite{asay-davis2016ExperimentalDesignThree,yung2026ResultsSecondIce}.
The ISOMIP+ configuration examines the response of the ice-shelf to far field ocean properties. The key difference between ISOMIP+ and our set-up is that we include surface salt forcing at the front of the ice shelf to mimic the formation of sea ice in coastal polynyas and consider a simpler cavity geometry with constant ice-ocean interface slope and a flat seabed.

The paper is organized as follows. In Section 2, we describe the ocean model set-up, the spatial domain, and the forcing scheme. 
In section 3, we demonstrate that our setup captures both the warm and cold circulation modes.
First, we characterize the steady response of the ice-shelf cavity to polynya forcing using quasi-static experiments and demonstrate that our configuration exhibits irreversible dynamics.
Then, we propose scaling laws for the basal melt rates in the warm and cold regimes and discuss physics-based predictions for the transition thresholds.
In Section 4, we present additional experiments exhibiting both irreversible and reversible transitions and demonstrate that the nature of regime changes (i.e. the irreversibility) is sensitive to far-field density stratification, ocean circulation at the front of the ice shelf, and the choice of subgrid-scale dissipation in the horizontal directions.
Finally, in section 5, we discuss the results in light of both recent observations and realistic simulations and highlight model limitations.

\section{Methods}

\subsection{Model description}

We simulate ocean flows within an ice-shelf cavity with the ocean component of the Nucleus for European Modelling of the Ocean (NEMO) model~\cite{madec2024nemo}. We use version 4.2.1 of the code.
The model solves the hydrostatic Boussinesq equations with the finite-difference method. The configuration is built upon the ISOMIP+ ice-shelf-ocean model intercomparison experiment~\cite{asay-davis2016ExperimentalDesignThree,yung2026ResultsSecondIce}. It includes the interactive ice-shelf NEMO module and uses z* vertical coordinates and partial cells~\cite{mathiot2017ExplicitRepresentationParametrised}.
The grid is uniform with a resolution step of 2 km horizontally and 20 m vertically, similarly to the ISOMIP+ setup~\cite{asay-davis2016ExperimentalDesignThree}. The time step is constant and equal to $360 \, \mathrm{s}$.
Non-diffusive processes are time-stepped with a Leapfrog scheme, and both tracer and momentum advections are approximated with second-order centered schemes.
The equations are solved on an $f$-plane ($f=-1.4\times10^{-4}$ $\mathrm{s^{-1}}$, corresponding to 75°S latitude).
No-slip lateral boundary conditions are applied on the sidewalls and a quadratic drag with dimensionless coefficient $C_d=2.5\times10^{-3}$ is applied on the seabed and underneath the ice shelf. The top surface of the open ocean is a rigid lid, consistent with the z*-coordinate framework.
The equation of state is linearized 
with thermal expansion coefficient $\alpha=3.733\times10^{-5}$ °C$^{-1}$ and haline contraction coefficient $\beta=7.843\times10^{-4}$ $\mathrm{(\psu)^{-1}}$. 
In order to prevent grid-scale noise induced by convective patterns in hydrostatic models~\cite{garreau2021ModelisationsObservationsContribution}, a biharmonic lateral diffusivity is implemented for both momentum and tracers with coefficients $\nu_H=\kappa_H=6.4\times10^7$ $\mathrm{m^4 \, s^{-1}}$. 
We note that the grid-scale noise could also have been suppressed using a third-order upwind advection scheme, which has numerical biharmonic diffusion.
Convective patterns are expected to develop when brine is released at the surface of the polynya. An Enhanced Vertical Diffusion scheme (EVD) is thus used to parameterize convective events. The EVD scheme works as follows: the Laplacian vertical diffusivity is $\nu_v=10^{-3}$ $\mathrm{m^2\,s^{-1}}$ for the momentum and $\kappa_v=5\times10^{-5}$ $\mathrm{m^2 \, s^{-1}}$ for the tracers if the stratification is stable between two cells; if the stratification becomes unstable, these values increase to $\nu_v^c=\kappa_v^c=0.1$ $\mathrm{m^2 \, s^{-1}}$ to model convection. 
All the model parameters and their associated standard values are summarized in~\ref{tab:parameter}.

\subsection{Spatial domain and forcing}

\begin{figure}[ht!]
\centering
\noindent\includegraphics[width=0.9\textwidth]{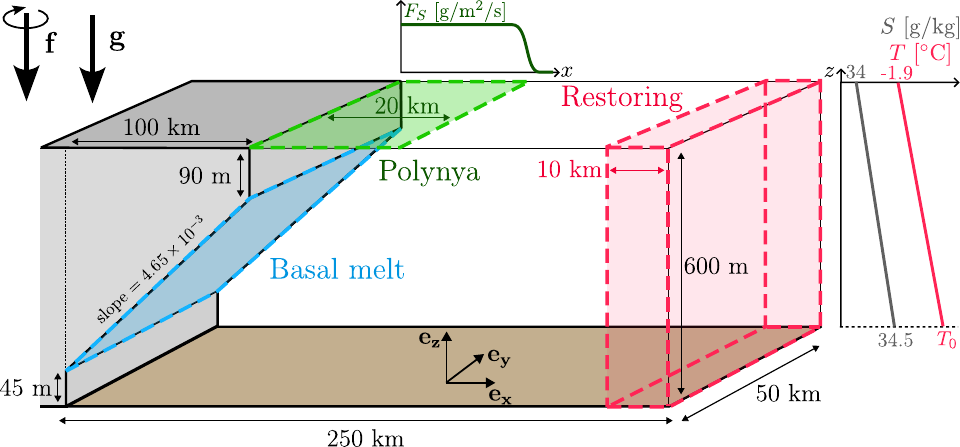}
\caption{Sketch of the idealized ocean ice-shelf domain. The ocean is bounded between the ice shelf, atmosphere and seabed. The interactive ice-ocean boundary (basal melt), far-field region (temperature and salinity restoring) and coastal polynya area (surface salt fluxes $F_S$ mimicking sea-ice formation) are highlighted in blue, red and green, respectively.}
\label{Fig1}
\end{figure}

The spatial domain is sketched in Fig.~\ref{Fig1}. The meridional extent ($x$-axis, ISOMIP+ convention) is 250 km and the zonal extent ($y$-axis) is 50 km. The seabed is flat, at depth 600 m. The ice shelf is 100 km long, 50 km large and has a fixed geometry. The ice-ocean interface has an idealized linear shape, as in~\cite{holland2008ResponseIceShelf,holland2017TransientResponseIce,hunter2006SpecificationTestModels}. Vertical walls are set at the grounding line and at the front of the ice shelf.

The sloped ice-shelf bottom boundary is interactive (in blue in Fig.~\ref{Fig1}) with basal melt computed and enforced along horizontal surfaces of the ice-ocean interface (not vertical), which has a discrete staircase pattern in the simulations. The ice-ocean interfacial fluxes are modeled with the three-equation model~\cite{holland1999ModelingThermodynamicIce}, neglecting the diffusion of heat through the ice shelf and ice salinity. The three equations are the linearized liquidus of the salt-water binary, the heat conservation equation and the salt conservation equation. They yield the temperature $T_b$ and salinity $S_b$ at the ice-ocean interface as well as the freshwater flux $q_\mathrm{fw}$ [$\mathrm{kg\,m^{-2}\,s^{-1}}$] (positive for melting and negative for freezing) from ocean properties within the ice-ocean boundary layer (also known as the Losh layer, ~\cite{losch2008ModelingIceShelf}), denoted by subscript ``$\mathrm{tbl}$'', as
\begin{linenomath}
\begin{subequations}
\begin{align}
T_b = \lambda_1 S_b + \lambda_2 + \lambda_3 H_\mathrm{isf}, \label{Eq:liquidus} \\
c_p \rho_\mathrm{tbl} \gamma_T (T_\mathrm{tbl}-T_b) = -\mathcal{L} q_\mathrm{fw}, \label{Eq:heat} \\
\rho_\mathrm{tbl} \gamma_S (S_\mathrm{tbl}-S_b) = -S_b q_\mathrm{fw}. \label{Eq:salinity}
\end{align}
\end{subequations}
\end{linenomath}
The temperature $T_\mathrm{tbl}$, salinity $S_\mathrm{tbl}$, and velocities $\mathbf{u_\mathrm{tbl}}$ are averaged (vertically) across the boundary layer, whose width is set to $h_\mathrm{tbl}=30$ m; $H_\mathrm{isf}$ [m] is the ice draft and $\lambda_1$ [$\mathrm{^\circ C\,(g\,kg^{-1})^{-1}}$], $\lambda_2$ [$\mathrm{^\circ C}$] and $\lambda_3$ [$\mathrm{^\circ C\,m^{-1}}$] are liquidus linearization coefficients given by~\cite{asay-davis2016ExperimentalDesignThree}, $c_p$ [$\mathrm{J\,kg^{-1}\,^\circ C^{-1}}$] is the ocean specific heat, and $\mathcal{L}$ [$\mathrm{J.kg^{-1}}$] is the ice latent heat of fusion given in~\cite{madec2024nemo}.
Importantly, $\gamma_T$ and $\gamma_S$ $\mathrm{[m\,s^{-1}]}$ are the turbulent heat and salt transfer velocities. They depend on the friction velocity $u_\ast$ in the top boundary layer such that $\gamma_{T,S}=\Gamma_{T,S} u_\ast$~\cite{jenkins2010ObservationParameterizationAblation}, where $u_\ast=\sqrt{C_d \left( \mathbf{u_\mathrm{tbl}}^2 + U_\mathrm{tide}^2 \right)}$, with $U_\mathrm{tide}=0.01$ m/s, a constant tidal velocity ensuring that $\gamma_T$ and $\gamma_S$ are always greater than 0. Here, $\Gamma_T=2.15\times10^{-2}$ and $\Gamma_S=6.14\times10^{-4}$ such that the ratio $\Gamma_T/\Gamma_S=35$, and the Stanton number is $\mathrm{St}=\Gamma_T \sqrt{C_d}=1.1\times10^{-3}$, as recommended by~\cite{jenkins2010ObservationParameterizationAblation}.
Evaporation is implemented uniformly at the sea surface to balance exactly the basal melt to prevent drifts of the sea-surface height~\cite{asay-davis2016ExperimentalDesignThree}; corrections of surface heat and salt fluxes are added simultaneously to avoid any surface buoyancy flux due to evaporation~\cite{mathiot2017ExplicitRepresentationParametrised}.

Brine rejection due to sea-ice formation within the polynya is modeled as a surface salt flux $F_S \, [\sfx]$ at the front of the ice shelf (highlighted by green area in Fig.~\ref{Fig1}), with a smooth transition at the edge of the polynya along the $x$-axis such that $F_S(x)=F_{S,0}\frac{\Theta(x-100\,\mathrm{km})}{2}\left\{1-\tanh\left[\frac{x-120\,\mathrm{km}}{3\,\mathrm{km}}\right]\right\}$ \cite{arthun2013EddyDrivenExchangeOpen}, with $\Theta$ the Heaviside step function.
We remark that the surface salt flux is the control parameter in our study, rather than the sea-ice formation rate, which is more readily available from observational studies~\cite{nakata2021MappingActiveFrazil}, because it represents the actual forcing in the simulations. Nonetheless, sea-ice formation rates $\xi$ linked to a prescribed $F_s$ value can be easily approximated with $\xi\,\mathrm{[m\,y^{-1}]}\approx1000\times F_s$, considering constant a sea-surface salinity of $34.2\psu$.

We model the far-field ocean as an infinite reservoir of seawater with prescribed temperature and salinity stratification using a finite-width restoring region where temperature and salinity are forced to relax towards imposed linear temperature and salinity profiles (cf. red region in Fig.~\ref{Fig1}). The restoring frequency $f_\mathrm{resto}$ is $0.1 \, \mathrm{days^{-1}}$ at the northern boundary and decays linearly to zero over 10 km in the southern direction~\cite{asay-davis2016ExperimentalDesignThree}. We did not see any qualitative changes of the overturning circulation considering 10 times higher or lower value for the restoring frequency.

Finally, a heat flux is imposed throughout the surface of the open ocean $Q_\mathrm{surf}=K(\mathrm{SST}-T_s)$ to restore the Sea Surface Temperature (SST) towards the surface freezing temperature $T_s=-1.9\,\mathrm{^\circ C}$ with a restoring coefficient $K=-40\,\mathrm{W\,m^{-2}\,^\circ C^{-1}}$.

\section{Results}

\subsection{Warm and cold circulation regimes} \label{subsec:2modes}

\begin{figure}[ht!]
\noindent\includegraphics[width=1.\textwidth]{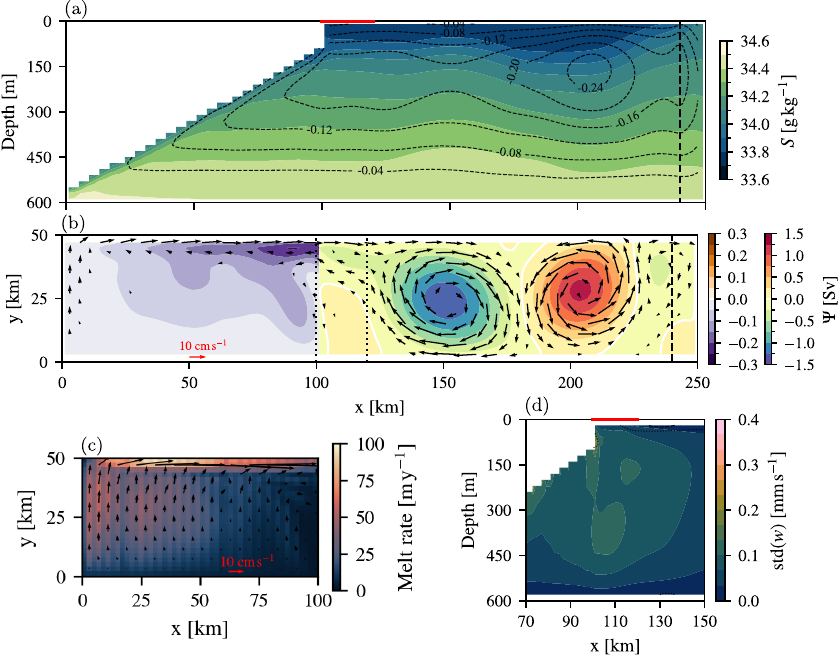}
\caption{Time averages of the last three years of the warm mode experiment. 
(a) Zonal average of the salinity field. The black contours show the Meridional Overturning Circulation (MOC) in Sverdrup (1 Sv = $10^6$ $\mathrm{m^3 \, s^{-1}}$). Solid lines denote anticlockwise rotation while dashed lines code for clockwise rotation. The red line highlights the polynya region.
(b) Barotropic stream function $\Psi$. Two color bars are used to emphasize the different magnitudes between the cavity and the open ocean. Black arrows represent the depth-averaged horizontal velocity (plotted only if greater than 2 $\mathrm{cm \, s^{-1}}$).
(c) Basal melt rate. Black arrows represent the horizontal velocity $\mathbf{u_\mathrm{tbl}}$ in the top boundary layer.
(d) Zonal mean of the standard deviation of the vertical velocity $w$. The dotted contour represents the spatial threshold for which the EVD scheme is active 10\% of the time. Note that the melt rate pattern in $x$ at the grid scale visible in (c) is a common artifact in cavity simulations using partial cells and with a staircase representation of the ice base~\cite{losch2008ModelingIceShelf}.}
\label{Fig2}
\end{figure}

\begin{figure}[ht!]
\noindent\includegraphics[width=1.\textwidth]{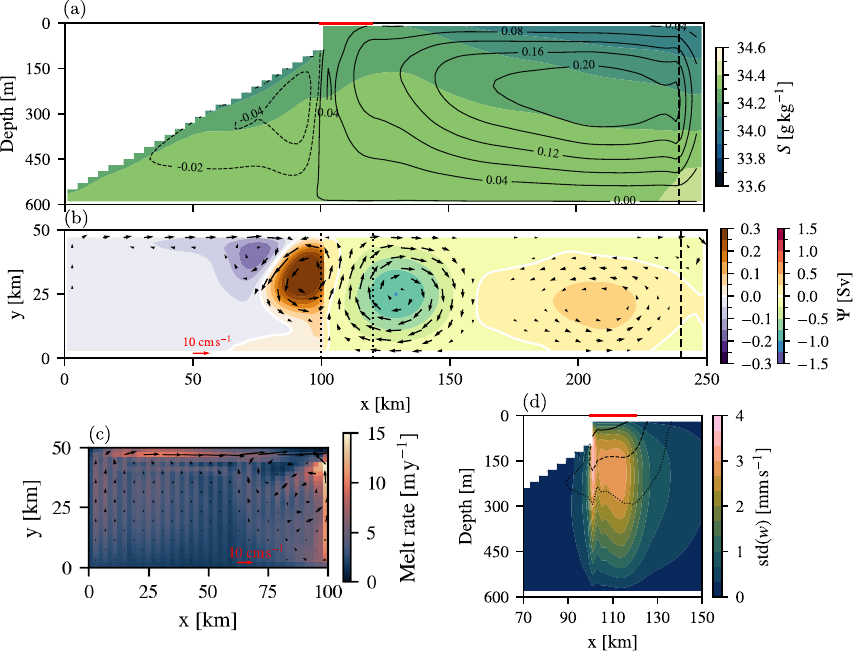}
\caption{Same as Fig.~\ref{Fig2}, but for the cold mode experiment (Surface salt flux $F_S$ is now set to $0.075\sfx$). Note that the scales of the color bars have changed in (c) and (d).  
In (d), the black contours represent the spatial thresholds for which the EVD scheme is active 10\% (dotted), 30\% (dashed), and 90\% (solid) of the time.}
\label{Fig3}
\end{figure}

We first test the ability of our setup to capture both the warm and the cold circulation regimes. For identical far-field ocean forcing, the surface salt forcing in the coastal polynya is expected to control the type of circulation in the ocean cavity~\cite{silvano2016OceanIceShelfInteraction}.
As a first step, we report the results of two experiments using very different surface forcing to clearly distinguish the warm and cold regimes. 
Specifically, we run two experiments wherein we impose the same restoring conditions in the far field but use different surface salt forcing: a warm-mode experiment without surface fluxes ($F_S=0$) and a cold-mode experiment with large surface salt fluxes $F_S=0.075 \sfx$. We note that the cold experiment considers extreme surface salt fluxes as extremal annual values observed around Antarctica range from zero when no sea ice is formed, especially beneath landfast sea ice~\cite{nakata2021MappingActiveFrazil}, to $0.03 \sfx$ in Cape Darnley, which is one of the most active sea-ice production region~\cite{ohshima2022DominantFrazilIce}. Linear restoring temperature and salinity profiles are imposed in the far field, ranging from $-1.9$ $^\circ$C and $34.0 \psu$ at the sea surface, to $1$ $^\circ$C and $34.5 \psu$ at the seafloor, which fall within the observed range of CDW properties. The resulting density profile is stable because the stratified salt profile, which stabilizes the water column, overcompensates the destabilizing stratified temperature field. The density ratio in the restoring region is $R_\rho=\frac{\beta \Delta S}{\alpha \Delta T}=3.6>1$ and the associated Brunt-Väisälä frequency is $N=2.2\times10^{-3}$ s$^{-1}$.
Both the warm- and cold-mode experiments are initialized with a quiescent ocean (zero velocities) and horizontally-uniform temperature and salinity profiles with the same vertical stratification as in the restoring region. We run the simulations for 10 years. Time averages of the last three years, which have reached statistical equilibrium, of each simulation are shown in Fig.~\ref{Fig2} for the warm mode experiment and in Fig.~\ref{Fig3} for the cold mode experiment.

The warm mode experiment reproduces well the Mode II described by~\cite{silvano2016OceanIceShelfInteraction} and shows strong similarity with the results of \citeA{holland2008ResponseIceShelf} and \citeA{holland2017TransientResponseIce} who used similar forcings.
In the remainder of this article, we will refer to this so-called Mode II as the ``warm regime" .
The ocean is strongly stratified throughout and a fresh water layer ($<34 \psu$), fed by significant meltwater rejections, is seen everywhere at the surface (Fig.~\ref{Fig2}(a)). The Meridional Overturning Circulation (MOC), defined as $\mathrm{MOC}=\int_{z'=-H}^z \int_{y=0}^{L_y}u\mathrm{d}z'\mathrm{d}y$, shows a single clockwise overturning cell, bringing warm waters from the deep part of the restoring region towards the grounding line of the ice shelf (Fig.~\ref{Fig2}(a)). The intense melting of the ice base sustains and reinforces the overturning cell as the upwelling  of relatively fresh waters along the ice-shelf slope pumps CDW into the cavity~\cite{jourdain2017OceanCirculationSeaice}.
The barotropic stream function, defined as $\Psi=\int_{x'=0}^x \int_{z=-H}^{0}v\mathrm{d}x'\mathrm{d}z$, shows the alternance of cyclonic and anti-cyclonic gyres along the channel, starting from the front of the ice shelf (Fig.~\ref{Fig2}(b)). Their extent is comparable to the width of the channel. Similar standing gyres have been observed within Greenland fjords adjacent to marine-terminating glaciers and described as resulting from the interplay between vorticity introduced by the upwelling meltwater plume and vorticity generated through friction along the fjord's topography~\cite{zhao2023StandingEddiesGlacial}.
Under the ice shelf, a cyclonic circulation is sustained by the upwelling meltwater that deviates towards the West due to the Coriolis force~\cite{hewitt2020SubglacialPlumes} and feed the cyclonic gyre at the ice-shelf front. The western intensification of the ice-shelf outflow, which is responsible for the high basal melt rates on the western side of the cavity (Fig.~\ref{Fig2}(c)), is consistent with the topographic beta effect due to the meridional variation of the water-column thickness. As expected, there are no convective patterns under the polynya ($F_s=0$) and the mean (not shown) and standard deviation of the vertical velocity are weak (cf. Fig.~\ref{Fig2}(d)). The EVD convective scheme is rarely active and only close to the sea surface, due to the surface heat flux temperature restoring (cooling).

In the cold mode experiment, the water column is less stratified in salinity, and therefore in density, in comparison with the warm mode experiment (Fig.~\ref{Fig3}(a)). Surface waters are only moderately fresh, suggesting weak meltwater rejections.
Instead of a single clockwise overturning circulation cell, two counter-rotating overturning cells are observed (Fig.~\ref{Fig3}(a)). On the one hand, we observe a clockwise overturning cell in the ice-shelf cavity, as in the warm mode experiment, although its strength is reduced because of the relatively low meltwater production. On the other hand, an anti-clockwise overturning cell is visible in the open ocean, with net downward volume fluxes from the sea surface towards the bottom of the ocean under the polynya, suggesting Dense Shelf Water (DSW) invades the cavity instead of Circumpolar Deep Water (CDW), consistent with the Mode I description~\cite{silvano2016OceanIceShelfInteraction}.
The barotropic stream function shows a succession of cyclonic and anti-cyclonic gyres along the channel as in the warm mode experiment, although their strengths are reduced because of the relatively weak meltwater outflow on the western edge of the channel (Fig.~\ref{Fig3}(b)). A gyre that only exists in the cold-mode experiment appears straddling the ice-shelf front. It is driven by the densification of the ocean under the polynya due to the brine releases, resulting in a thermal-wind anticyclonic circulation~\cite{arthun2013EddyDrivenExchangeOpen}.
The western boundary current still exists but is relatively weak since DSW is cold and drives low basal melt rates.
The anti-cyclonic gyre straddling the ice-shelf front allows surface waters to intrude the front of the ice-shelf cavity with high velocities, enhancing local melt rates in a fashion reminiscent of the mode III melting regime that is sometimes observed along the front of ice shelves~\cite{silvano2016OceanIceShelfInteraction}.
In the remainder of the article, the name ``cold regime" will thus denote a combination of the circulation Mode I and III.
Strong vertical velocity signals are observed under the polynya, as expected for regions with deep convection (Fig.~\ref{Fig3}(d)). The zonally averaged standard deviation of the vertical velocity is large everywhere under the polynya, as we partially resolve convective chimneys~\cite{xu2023PartI}. The mean of the vertical velocity (not shown) is one order of magnitude weaker than the standard deviation. Thus, vertical mixing is the predominant mechanism modifying the deep waters properties, rather than a net downward volume flux of cold DSW.
Fig.~\ref{Fig3}(d) shows that the EVD scheme is frequently active close to the surface of the polynya, where brine rejections destabilize the water column.

\subsection{Hysteresis experiments} \label{subsec:hysteresis}

\begin{table}
\caption{Summary of the quasi-static numerical experiments used to investigate regime transitions and hysteresis. TEMP1 is the reference configuration for the sensitivity experiments, and parameters varied relative to TEMP1 are highlighted in bold. Experiments are grouped according to variations in far-field temperature (TEMP), salinity stratification (SSS and SBS), lateral boundary conditions (FSLIP), and horizontal viscosity (NUH). SSS, SBS, and LBC denote restoring Sea Surface Salinity, restoring Sea Bottom Salinity, and Lateral Boundary Condition, respectively.}
\centering
\begin{tabular}{l|c c c c c c}
 & $\mathbf{T_0}$ & $\mathbf{F_S}$ \textbf{range} & \textbf{SSS} & \textbf{SBS} & $\mathbf{\nu_H}$ & \textbf{LBC} \\
\textbf{EXP}  & $[^\circ C]$ & $[\sfx]$ & [$\psu$] & $[\psu]$ & $[\times10^7\mathrm{m^4\,s^{-1}}]$ & [-]  \\
\hline
TEMP-1.9 & $\mathbf{-1.9}$  & $[0,0.075]$ & $34$   & $34.5$ & $6.4$   & No slip \\
TEMP-1   & $\mathbf{-1}$    & $[0,0.075]$ & $34$   & $34.5$ & $6.4$   & No slip \\
TEMP0    & $\mathbf{0}$     & $[0,0.075]$ & $34$   & $34.5$ & $6.4$   & No slip \\
TEMP1    & $1$     & $[0,0.075]$   & $34$   & $34.5$ & $6.4$   & No slip \\
TEMP2    & $\mathbf{2}$     & $[0,0.075]$ & $34$   & $34.5$ & $6.4$   & No slip \\
TEMP0.5  & $\mathbf{0.5}$   & $[0,0.04]$  & $34$   & $34.5$ & $6.4$   & No slip \\
TEMP1.5  & $\mathbf{1.5}$   & $[0,0.06]$  & $34$   & $34.5$ & $6.4$   & No slip \\
\hline
SSS33.8  & $1$     & $[0,0.075]$ & $\mathbf{33.8}$ & $34.5$ & $6.4$   & No slip \\
SSS34.2  & $1$     & $[0,0.075]$ & $\mathbf{34.2}$ & $34.5$ & $6.4$   & No slip \\
SBS34.7  & $1$     & $[0,0.075]$ & $34$   & $\mathbf{34.7}$ & $6.4$   & No slip \\
SBS34.3  & $1$     & $[0,0.075]$ & $34$   & $\mathbf{34.3}$ & $6.4$   & No slip \\
\hline
FSLIP    & $1$     & $[0,0.1]$   & $34$   & $34.5$ & $6.4$   & \textbf{Free slip} \\
\hline
NUH1.3   & $1$     & $[0,0.075]$   & $34$   & $34.5$ & $\mathbf{1.3}$ & No slip \\
NUH32    & $1$     & $[0,0.075]$   & $34$   & $34.5$ & $\mathbf{32}$  & No slip 
\end{tabular}
\label{table:exp}
\end{table}

We now investigate the nature and reversibility of the transitions between the cold and warm circulation regimes. To this end, we perform quasi-static experiments in which the surface salt flux imposed $F_s$ is varied stepwise. By comparing simulations in which $F_s$ is increased or decreased, we determine whether the transition occurs at the same forcing threshold in both directions or instead exhibits hysteresis.

We evaluate how the far-field ocean temperature influences regime transitions using five different restoring temperature profiles, corresponding to experiments TEMP-1.9 to TEMP2 in Table~\ref{table:exp} (cf. first five lines). All profiles have the same surface temperature $T_s=-1.9\cels$ but different bottom temperatures, i.e. $T_0\in\{-1.9,-1,0,1,2\cels\} $. In each experiment, the restoring temperature varies linearly with depth between $T_s$ and $T_0$. The salinity restoring profile is kept identical in all cases and increases linearly from $34\psu$ at the surface to $34.5\psu$ at the seafloor. This experimental design isolates the influence of far-field temperature on the transition while maintaining the same background salinity stratification.
For each far-field temperature profile, we perform two simulations following opposite forcing trajectories. The increasing-$F_s$ branch is designed to probe the transition from the warm to the cold regime. The ocean is initially set to the temperature and salinity profiles prescribed in the restoring region, which favors the warm circulation regime when such a regime exists. Starting from $F_s=0\sfx$, the surface salt flux is then increased by $0.005\sfx$ every 10 years until it reaches $F_s=0.075\sfx$.
The decreasing-$F_s$ branch is instead designed to probe the transition from the cold to the warm regime. To initialize the system in the cold regime, the salinity is again set to the vertically varying restoring profile, while the temperature is initialized uniformly at $-1.9\cels$. Starting from $F_s=0.075\sfx$, the surface salt flux is then decreased by $0.005\sfx$ every 10 years until $F_s=0\sfx$ is reached.

Each branch contains 16 forcing levels and spans 160 years. At a given forcing level, the main circulation properties generally equilibrate within the first few years of the 10-year interval. Therefore, we characterize the quasi-steady response of the system using averages over the final three years of each interval. Comparing the two forcing branches then allows us to identify both the transition thresholds and the range of $F_s$ over which the cold and warm regimes coexist.

Note that the warm and cold circulation regimes described in the previous section (Figs.~\ref{Fig2} and~\ref{Fig3}) correspond to the two end-member states of the reference TEMP1 experiment (Table~\ref{table:exp}).

\begin{figure}[ht!]
\centering
\noindent\includegraphics[width=1.\textwidth]{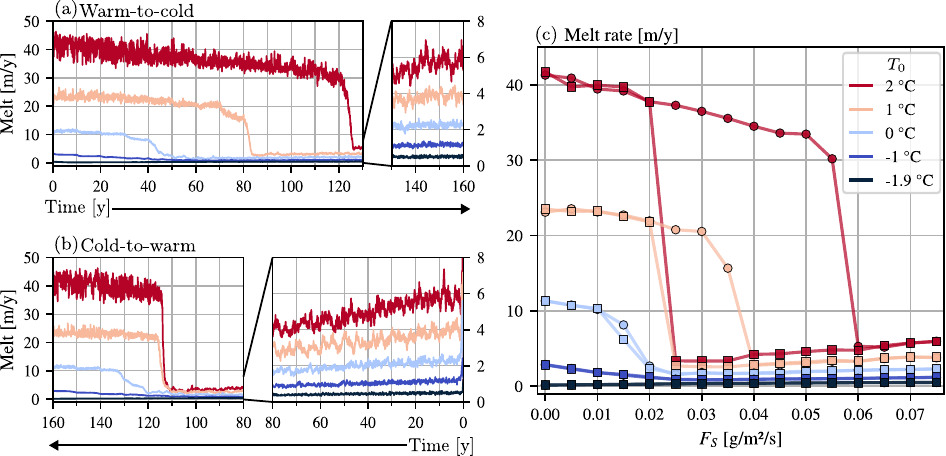}
\caption{(a) Spatially and monthly averaged basal melt rates for the warm-to-cold experiments, plotted as functions of time. The surface salt flux $F_S$ increases from $0$ to $0.075 \sfx$ in $0.005 \sfx$ increments every 10 years.
(b) Same as (a), but for the cold-to-warm experiments, with $F_S$ decreasing from $0.075$ to $0 \sfx$ also with $0.005 \sfx$ increments every 10 years. 
(c) Basal melt rates averaged over space and over the final 3 years of each 10-year window plotted against the imposed surface salt flux. Results from the warm-to-cold (resp. cold-to-warm) experiments are shown with circles (resp. squares).}
\label{Fig4}
\end{figure}

Fig.~\ref{Fig4}(a) and Fig.~\ref{Fig4}(b) show the temporal evolution of the spatially and monthly averaged basal melt rates of the warm-to-cold and cold-to-warm experiments, respectively.
The time axis of the cold-to-warm experiment (bottom plot) has been inverted such that $F_S$ increases from left to right, i.e. as for the warm-to-cold case (top plot).
The behavior of the melt rate time series depends on the bottom temperature in the restoring zone: for the coldest temperatures ($-1\,^\circ$C and $-1.9\,^\circ$C), the melt rate does not vary much (dark blue curves). As the bottom temperature increases, we start observing  transitions between high and low melt rate regimes. The transition becomes more abrupt as the bottom temperature in the restoring zone increases: it occurs within a few years for the $1\,^\circ$C and $2\,^\circ$C temperature forcing (see, e.g., the dark red curve around year 125 in Fig.~\ref{Fig4}(a)).
We note that the system is close to having relaxed to a new steady state by the end of each 10-year segment of the hysteresis experiment for most forcing increments.
We show in Fig.~\ref{Fig4}(c) the basal melt rates averaged over the final 3 years of each 10-year increment, as a function of the imposed surface salt flux. Time averaging removes the internal variability and more clearly shows the steady response of the ice-shelf to the coastal polynya forcing.
Fig.~\ref{Fig4}(c) highlights that the salt flux thresholds for the abrupt transitions ($1\,^\circ$C and $2\,^\circ$C) are not the same for the increasing (circles) and decreasing (squares) $F_S$ branches of the experiment. The basal melt rates then exhibit a hysteresis behavior as a function of $F_S$ and regime transitions are irreversible.
Reversible transitions are recovered for low forcing temperatures ($T_0\in\{0,-1,-1.9\}\cels$), as can be seen from the overlap of the blue circles and squares over the full range of $F_s$ values tested.

\begin{figure}[ht!]
\centering
\noindent\includegraphics[width=1.\textwidth]{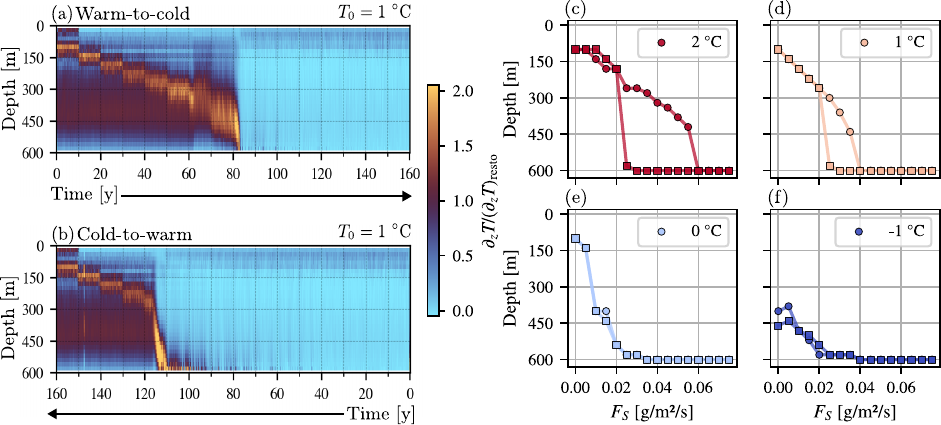}
\caption{(a,b) Ratio between the temperature stratification averaged over the polynya region and every month, and the temperature stratification imposed in the restoring region (TEMP1, $T_0=1\cels$), plotted as a function of time, for the warm-to-cold experiment in (a), and the cold-to-warm experiment in (b).
The time axis of the cold-to-warm experiment has been inverted to show increasing values of $F_S$ from the left to the right in both plots.
(c-f) Thermocline depth (depth of the maximal value for $\partial_zT/\partial_z T_\mathrm{resto}$) averaged over the polynya region and over the final 3 years of each 10-year window for warm-to-cold experiments (circle) and cold-to-warm experiments (squares).}
\label{Fig5}
\end{figure}

The extent of the bistable region in terms of surface salt fluxes increases with the melt rate in the warm regime (visually, taller and wider hysteresis). Thus, we postulate that the meltwater feedback in our simulations works such that the thick meltwater layer in the warm regime delays deep convection in the polynya region.
To verify this hypothesis, we show in Fig.~\ref{Fig5}(a) and Fig.~\ref{Fig5}(b) the temporal evolution of the temperature gradient $\partial_zT$ as a function of depth below the polynya, normalized by the temperature gradient $\partial_zT_\mathrm{resto}$ in the restoring region, for the warm-to-cold and cold-to-warm experiments ($T_0=1\,^\circ$C), respectively.
Mixed conditions are highlighted by light blue colors ($\partial_zT/\partial_zT_\mathrm{resto}\ll1$), stratified conditions similar to in the restoring region are highlight by dark blue colors ($\partial_zT/\partial_zT_\mathrm{resto}\approx1$), and strongly-stratified conditions (denoting the presence of a thermocline) are shown in yellow ($\partial_zT/\partial_zT_\mathrm{resto}>1$). The mixed layer is fed by DSW, whereas the stratified layer below the thermocline is supplied by CDW. Thus, the thermocline depth controls which water masses are intruding the cavity. 
In the warm mode, the thermocline does not reach the bottom of the ocean, and unmodified CDW can intrude the cavity (dark blue colors dominate).
In the cold mode, the water column is fully mixed as convective patterns reach the bottom of the ocean. Thus, DSW mix with CDW, resulting in colder waters entering the cavity.
The transition from high to low basal melt rates occurs when the mixed layer depth reaches the bottom of the ocean, i.e. at $t\approx 80$ years in both Fig.~\ref{Fig4}(a) (orange solid line) and Fig.~\ref{Fig5}(a). 
Conversely, the transition from low to high basal melt rates occurs when the thermocline rises from the bottom of the ocean, i.e. at $t\approx 110$ years (see Fig.~\ref{Fig4}(b) and Fig.~\ref{Fig5}(b)). 
Fig.~\ref{Fig5}(c-f) show the depth of the maximum  of the temperature gradient averaged over the final 3 years of each 10-year increment (i.e., the depth of the thermocline ) as a function of the imposed surface salt flux.
Note that if the maximum temperature gradient is inferior to the temperature gradient in the restoring region, we consider that deep convection occurs and set the thermocline to the bottom depth.
As previously seen for basal melting, a hysteresis dynamics is observed when the ocean is forced with warm enough waters ($T_0\in\{1,2\}\cels$), whereas continuous transitions are observed in colder experiments ($T_0\in\{0,-1\}\cels$).
For irreversible transitions (Fig.~\ref{Fig5}(c,d)), starting in the warm mode, a linear increase in the depth of the mixed layer is observed first when $F_S$ increases, before it abruptly collapses toward the bottom of the ocean in the warm-to-cold transition. The variability of the thermocline depth increases as the system approaches the transition threshold, as shown in Fig.~\ref{Fig5}(a).
In contrast, in the cold mode, decreasing the value of $F_S$ does not affect the depth of the mixed layer, which stays fixed on the sea floor. At the cold-to-warm transition threshold, the thermocline abruptly leaves the seafloor, resulting in an abrupt restratification of the deep part of the ocean. Before the transition, episodic intrusion of warm deep waters are observed (larger values of $\partial_zT/\partial_z T_\mathrm{resto}$ close to the seafloor) and they become more frequent when approaching the transition threshold, as shown in Fig.~\ref{Fig5}(b).

\subsection{Scaling laws}

In this section we investigate the dependence of the basal melt rate with the control parameters and aim to predict the thresholds for regime transitions. The subsequent analysis is based on the results of the simulations discussed in Section~\ref{subsec:hysteresis} as well as the additional TEMP0.5 and TEMP1.5 experiments listed in Table~\ref{table:exp}, with $T_0=0.5\cels$ and $1.5\cels$, respectively.

\subsubsection{Basal melt rates scaling}

\begin{figure}[ht!]
\centering
\noindent\includegraphics[width=3.2 in]{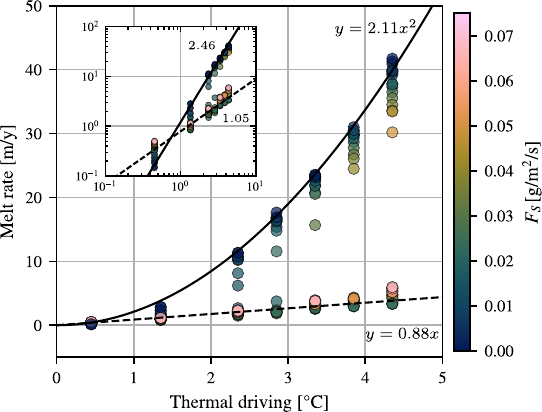}
\caption{Steady-state basal melt rates shown as a function of thermal driving. A quadratic fit is shown for simulation results in the warm mode with $F_s=0$ (solid line), and a linear fit is shown for simulation results in the cold mode with $F_s=0.04\sfx$ (dashed line).
The inset shows the same results but use logarithmic axes to highlight the comparison with the best-fit power laws.}
\label{FigScaleMelt}
\end{figure}

We first investigate the dependence of the melt rate  with thermal driving, which we define as the difference between the bottom temperature $T_0$ in the restoring region and the local freezing point $T_b(H)=-2.33 \cels$ evaluated with constant reference salinity $34.2 \psu$ (cf. Eq.~(\ref{Eq:liquidus})), i.e.  $T_0-T_b(H)$.
Fig.~\ref{FigScaleMelt} shows the spatial average of the basal melt rate as a function of thermal driving.
The melt rates corresponding to the warm and cold modes are reasonably well captured by two distinct curves.
The upper curve approximates melt rates in the warm mode with a quadratic equation $y=ax^2$ (solid line), consistent with~\cite{holland2008ResponseIceShelf}, which can be inferred from Eq.~\ref{Eq:heat}. In Eq.~\ref{Eq:heat}, the meltwater flux at the ice–ocean interface is proportional to the thermal forcing $T_\mathrm{tbl}-T_b$, and to the turbulent heat transfer coefficient $\gamma_T$. The coefficient $\gamma_T$ depends linearly on $\mathbf{u_\mathrm{tbl}}$, which according to meltwater plume theory, also varies linearly with thermal forcing~\cite{hewitt2020SubglacialPlumes}.
Thus, assuming a linear dependency between the thermal forcing and the thermal driving, the basal melt rates are expected to vary quadratically with thermal driving.
In the inset, a best-fit minimizing the error for the log of the melt rate yields a power law exponent of 2.46 for simulation results in the warm mode. This larger than 2 exponent could be explained because basal freezing occurs under the ice shelf when thermal driving is weak, resulting in an anomalously low spatial average of the basal melt rate at low temperatures, and a steeper slope for the power law fit. 
A power-law fit of basal melt rates in the warm mode with thermal driving values greater than $2\cels$ yields a 2.06 power law exponent.
On the other hand, we find that melt rates in the cold mode follow a quasi linear equation $y=ax$ (dashed line) ; a best-fit minimizing the logarithmic of the melt rate in the cold mode yields  a 1.05 power law exponent.
The fact that the quadratic melting law is not recovered in the cold mode is not surprising since water masses entering the cavity are strongly modified by surface waters when convection is active. Additionally, cold circulations exhibit a Mode III melting regime close to the ice-shelf front, which can be expected to further decrease the sensitivity of the melt rate to the far-field bottom temperature. Thus, scaling laws for thermal forcing and $\mathbf{u_\mathrm{tbl}}$ should also deviate from those obtained within the meltwater plume theory framework.

\begin{figure}[ht!]
\centering
\noindent\includegraphics[width=3.2 in]{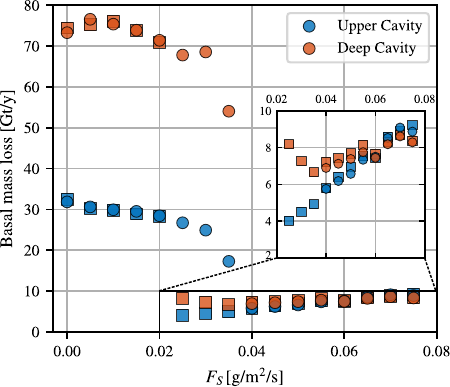}
\caption{Steady-state basal mass losses for the TEMP1 experiment ($T_0=1\cels$). Red (resp. blue) data points represent average basal mass losses in the deeper (resp. shallower) part of the cavity, using a 300 m depth threshold.
Warm-to-cold (resp. cold-to-warm) data points are shown by circles (resp. squares).}
\label{FigDeepUp}
\end{figure}

The basal melt rate dependence on $F_s$ is shown in Fig.~\ref{Fig4}. In addition to the sharp decrease obtained when the cavity shifts from warm to cold conditions, we find a decrease (resp. increase) of the basal melt rate with increasing $F_S$ in the warm mode (resp. cold mode). The two trends are opposite because they are controlled by two different types of dynamics, which impact different parts of the cavity.
Fig.~\ref{FigDeepUp} shows the basal mass loss contributions of the deeper part (ice draft $>300$ m) and the upper part (ice draft $<300$ m) of the cavity.
In the warm mode, the deeper part of the cavity, where thermal driving is high, is the primary contributor to meltwater formation. Meltwater masses formed at depth cool as they upwell along the ice shelf, leading to relatively weak basal melt rates in the upper part of the cavity, as predicted by meltwater plume models~\cite{lazeroms2018ModellingPresentdayBasal,hewitt2020SubglacialPlumes}.
The decrease of basal melt rates with increasing $F_S$ in the warm mode is then due to the lowering of thermal driving at depth, which is driven by thermocline deepening (see Fig.~\ref{Fig5}) and deep water mixing with surface water under the polynya. 
In the cold mode, Fig.~\ref{FigDeepUp} shows that the basal mass loss in the deeper part of the cavity is weakly impacted by the surface salt fluxes, whereas the basal mass loss of the upper part increases with $F_S$ (cf. inset). 
In the deeper part of the cavity, basal melt rates are linked to the mode I circulation, which has deep water masses with surface freezing temperature. As the salinity impacts the freezing temperature rather weakly, the response of the deep basal mass loss to the surface salt fluxes is very small and nearly constant.
In the upper part of the cavity, basal melt rates are instead controlled by a mode III circulation, which features surface waters intruding below the ice-shelf front, with relatively weak thermal forcing but high top boundary layer velocities $\mathbf{u_\mathrm{tbl}}$ (see Fig.~\ref{Fig3}(c)).
Importantly, the thermal-wind balance that underpins the mode III circulation is driven by surface salt flux~\cite{arthun2013EddyDrivenExchangeOpen}.
Thus, the global basal melt increases with $F_S$ in the cold mode due to the increasing contribution from the Mode III circulation.

\subsubsection{Transition thresholds and bistability scaling}

The identification of regime thresholds in phase space is key to assessing how far a cavity stands from a regime shift and to quantifying irreversibility.
Previous low-dimensional studies suggested that a buoyancy (or freshwater) budget analysis could be used to predict the state of a cavity and the range of bistable dynamics~\cite{hazel2020BistabilityFilchnerRonneIce,moorman2023CoastalPolynyasEnable,finucane2026TheoryMeltwaterPolynya,saddier2026IrreversibleTransitionsOcean}.
According to these studies, the buoyancy flux budget (or trend) should be negative under the polynya in the convective cold mode, positive in the warm stably stratified mode, and at equilibrium at the transition threshold.
Here, the buoyancy flux $B_\mathrm{polynya}$ $\mathrm{[m^4\,s^{-3}]}$ associated with the imposed surface salt fluxes reads
\begin{linenomath}
\begin{equation}
B_\mathrm{polynya} = -\frac{\beta g}{\rho_\star} \mathcal{A}_p F_S , \label{Eq:Bsfx} 
\end{equation}
\end{linenomath}
where $\mathcal{A}_p$ is the polynya area and $g$ [$\mathrm{m/s^2}$] is the gravity acceleration constant.
Note that we neglected here the buoyancy loss associated to the surface heat restoring. 
The other buoyancy flux key to the ice-front polynya buoyancy budget, i.e. $B_\mathrm{melt}$ $\mathrm{[m^4/s^3]}$, is due to the meltwater fluxes, and can be expressed as
\begin{linenomath}
\begin{equation}
B_\mathrm{melt} = \frac{\beta g}{\rho_\star} \mathcal{A}_c \langle S_b q_\mathrm{fw} \rangle 
-\frac{\alpha g}{\rho_\star} \mathcal{A}_c \frac{\mathcal{L}}{c_p} \langle q_\mathrm{fw} \rangle, \label{Eq:Bmelt} 
\end{equation}
\end{linenomath}
where $\mathcal{A}_c$ is the horizontal area of the ice shelf, $\langle S_b q_\mathrm{fw} \rangle$ [$\sfx$] is the steady-state salt flux (brackets denote time averages) at the ice-ocean interface, and $\langle q_\mathrm{fw} \rangle$ [$\mathrm{kg \, m^{-2} \, s^{-1}}$] is the steady-state freshwater flux at the ice-ocean interface; both boundary fluxes being output of the three-equation model (Eqs.~\eqref{Eq:liquidus} to \eqref{Eq:salinity}).
Setting Eq.~\eqref{Eq:Bsfx} equal to Eq.~\eqref{Eq:Bmelt} yields a value for the surface salt flux at equilibrium, which depends on the (initial) circulation regime. Setting $B_\mathrm{polynya}=B_\mathrm{melt}$ for each of the two regimes thus yields two values for $F_s$ describing both regime transitions, i.e.
\begin{linenomath}
\begin{subequations}
\begin{align}
F_S^\mathrm{cold \to warm} &= \frac{\mathcal{A}_c}{\mathcal{A}_p} \left(\langle S_b q_\mathrm{fw}^\mathrm{cold} \rangle-\frac{\alpha\mathcal{L}}{\beta c_p}\langle q_\mathrm{fw}^\mathrm{cold} \rangle\right), \label{Eq:Fs_ctw} \\
F_S^\mathrm{warm \to cold} &=\frac{\mathcal{A}_c}{\mathcal{A}_p} \left(\langle S_b q_\mathrm{fw}^\mathrm{warm} \rangle-\frac{\alpha\mathcal{L}}{\beta c_p}\langle q_\mathrm{fw}^\mathrm{warm} \rangle\right) , \label{Eq:Fs_wtc}
\end{align}
\end{subequations}
\end{linenomath}
where $q_\mathrm{fw}^\mathrm{cold}$ (resp. $q_\mathrm{fw}^\mathrm{warm}$) is the freshwater flux just before the cold-to-warm (resp. warm-to-cold) transition.

\begin{figure}[ht!]
\centering
\noindent\includegraphics[width=3.2 in]{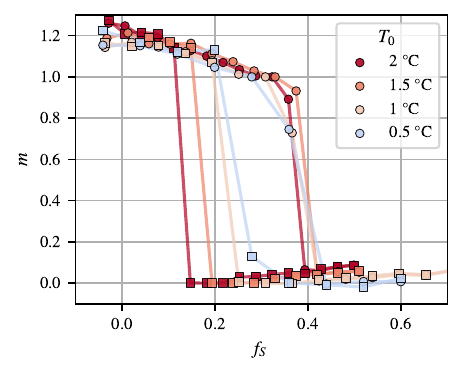}
\caption{Dimensionless melt water flux $m$ as a function of the dimensionless surface salt flux $f_S$.}
\label{FigHystSize}
\end{figure}

In Fig.~\ref{FigHystSize}, we plot a normalized version of the meltwater flux $m$ as a function of the normalized surface salt flux $f_S$, defined as
\begin{linenomath}
\begin{subequations}
\begin{align}
m&=\frac{\langle q_\mathrm{fw} \rangle - \langle q_\mathrm{fw}^\mathrm{cold} \rangle}{\langle q_\mathrm{fw}^\mathrm{warm} \rangle - \langle q_\mathrm{fw}^\mathrm{cold} \rangle}, \label{Eq:m} \\
f_S&=\frac{F_S-F_S^\mathrm{cold \to warm}}{F_S^\mathrm{warm \to cold}-F_S^\mathrm{cold \to warm}}, \label{Eq:fs}
\end{align}
\end{subequations}
\end{linenomath}
for all bistable experiments to test whether transition thresholds in dimensionless space always occur at $f_S=0$ (resp. $f_S=1$) for the cold-to-warm (resp. warm-to-cold) transition, as suggested by the buoyancy budget analysis. Note that $q_\mathrm{fw}^\mathrm{cold}$ and $q_\mathrm{fw}^\mathrm{warm}$ have been chosen as the second to last meltwater fluxes before each transition, because the values of the last points are more sensitive to their distance to the transition thresholds.
The dimensionless curves obtained are almost superimposed for all experiments, suggesting that the buoyancy budget analysis captures reasonably well the sensitivity of basal melt rates and transition thresholds to the far-field temperature.
The warm-to-cold transitions collapse on the same master curve but occur at $f_S\approx0.4$ instead of the predicted value $f_S=1$. This discrepancy is not surprising since the water column under the polynya is an open system that can have horizontally-diverging buoyancy fluxes when convection sets in. The lower than one $f_S$ threshold suggests that some of the meltwater produced in the cavity is advected away from the polynya region, as already shown by Xu et al.~\cite{xu2023PartI,xu2023PartII}, and is in good agreement with previous low-dimensional models of ocean cavities that adjusted Eqs.~\eqref{Eq:Fs_ctw} and \eqref{Eq:Fs_wtc} with a dimensionless coefficient inferior to one to match transition thresholds of numerical simulations~\cite{hazel2020BistabilityFilchnerRonneIce,finucane2026TheoryMeltwaterPolynya}.
The cold to warm transitions show slightly more scatter and occur for values superiors to the predicted threshold $f_S=0$. This suggests that restratification occurs before reaching an equilibrium buoyancy budget, possibly because of the advection of meltwater outside of the polynya region.

\section{From irreversible to reversible regime transitions}

We now examine the sensitivity of the hysteresis behavior using the SSS/SBS, FSLIP, and NUH experiments summarized in Table~\ref{table:exp}. More precisely, the SSS and SBS experiments modify the strength of the salt stratification in the restoring region, the FSLIP experiment impacts the large-scale flow structures at the ice-shelf front through reduced sidewall friction, and the NUH experiments modify the subgrid-scale dissipation scheme.

\subsection{Impact of salt stratification} \label{subsec:saltstrat}

\begin{figure}[ht!]
\centering
\noindent\includegraphics[width=1.\textwidth]{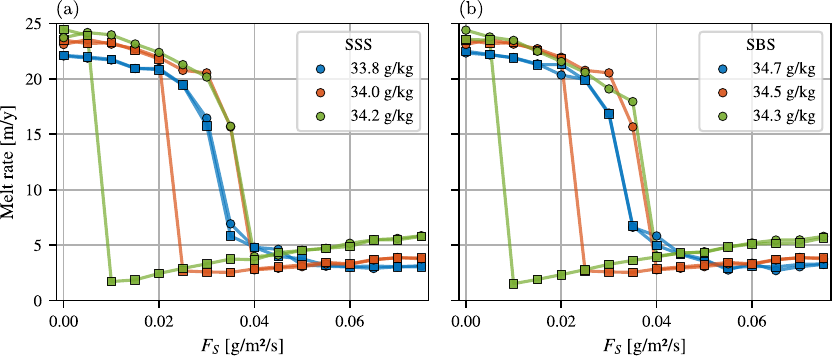}
\caption{Steady-state basal melt rates as a function of surface salt fluxes for $T_0=1\cels$ for different (a) Sea Surface Salinity (SSS) or (b) Sea Bottom Salinity (SBS) values imposed within the restoring region. Circles (resp. squares) denote results obtained with $F_s$ increasing (resp. decreasing). We set $\mathrm{SBS}=34.5\psu$ for the experiments shown in (a) and $\mathrm{SSS}=34.0\psu$ for those shown in (b), such that the green, orange and blue curves show experiments with the same weak, reference and strong stratification in both (a) and (b), respectively.}
\label{Fig6}
\end{figure}

We first test the impact of the salinity field in the restoring region. We restrict our attention to the case of moderate temperature forcing, i.e., $T_0=1\cels$ (cf. orange curve in Fig.~\ref{Fig4}). Fig.~\ref{Fig6}(a) shows the basal melt rates obtained for three quasi-static experiments with different Sea Surface Salinity (SSS) of $33.8\psu$, $34.0\psu$, and $34.2\psu$. These configurations correspond to experiments SSS33.8, TEMP1 (reference), and SSS34.2 in Table~\ref{table:exp}. The corresponding Brunt-Väisälä frequencies are $8.5\times10^{-3}\,\mathrm{s^{-1}}$, $2.2\times10^{-3}\,\mathrm{s^{-1}}$, and $1.4\times10^{-3}\,\mathrm{s^{-1}}$, respectively.
An increase of the density stratification in the restoring region results in an increase of the cold-to-warm transition threshold (leftmost bifurcation), diminishing the size of the bistable region. For the largest density stratification, the bistable region completely shrinks, and the transition becomes fully reversible.
The impact of the restoring density stratification on the cold-to-warm transition threshold can be understood with a simple argument: stronger density stratification in the restoring region favors restratification in the polynya region, and conversely.
In contrast, the warm-to-cold transition threshold is largely insensitive to changes in the restoring salinity stratification. This asymmetry suggests that the warm-to-cold transition is controlled more strongly by processes associated with basal melting and the resulting freshwater input near the ice-shelf front than by the far-field density stratification itself. 
The results shown in Fig.~\ref{Fig4} support this interpretation. The warm-to-cold transition threshold varies strongly with $T_0$, which directly controls the amount of basal melting and freshwater input in the warm regime. By contrast, varying $T_0$ has little effect on the cold-to-warm threshold, consistent with the fact that these experiments have similar far-field density stratifications. 
Taken together, these results suggest that the two transition thresholds are governed by distinct mechanisms: the cold-to-warm transition is primarily sensitive to the far-field density stratification, whereas the warm-to-cold transition is more strongly influenced by the meltwater releases at the ice-shelf front.

To test whether the Sea Bottom Salinity can similarly impact the dynamics as the Sea Surface Salinity in the restoring region, we show in Fig.~\ref{Fig6}(b) the results of three quasi-static experiments obtained with SBS set to $34.7\psu$, $34.5\psu$, and $34.3\psu$. These configurations correspond to experiments SBS34.7, TEMP1 (reference), and SBS34.3 in Table~\ref{table:exp}, respectively. The SSS is set equal to $34.0\psu$, such that the Brunt-Väisälä frequencies are the same as those imposed in the SSS experiments.
The experiments with different SBS values give the exact same basal melt rates as those obtained with different SSS values. Thus, the absolute values of the imposed salinities are not important for the dynamics and only the relative values of the imposed salinity profiles play a role, in agreement with recent results from low-dimensional models~\cite{hazel2020BistabilityFilchnerRonneIce,moorman2023CoastalPolynyasEnable,saddier2026IrreversibleTransitionsOcean}.

\subsection{Impact of the large scale circulation: from no-slip to free-slip lateral boundaries}

\begin{figure}[ht!]
\noindent\includegraphics[width=1.\textwidth]{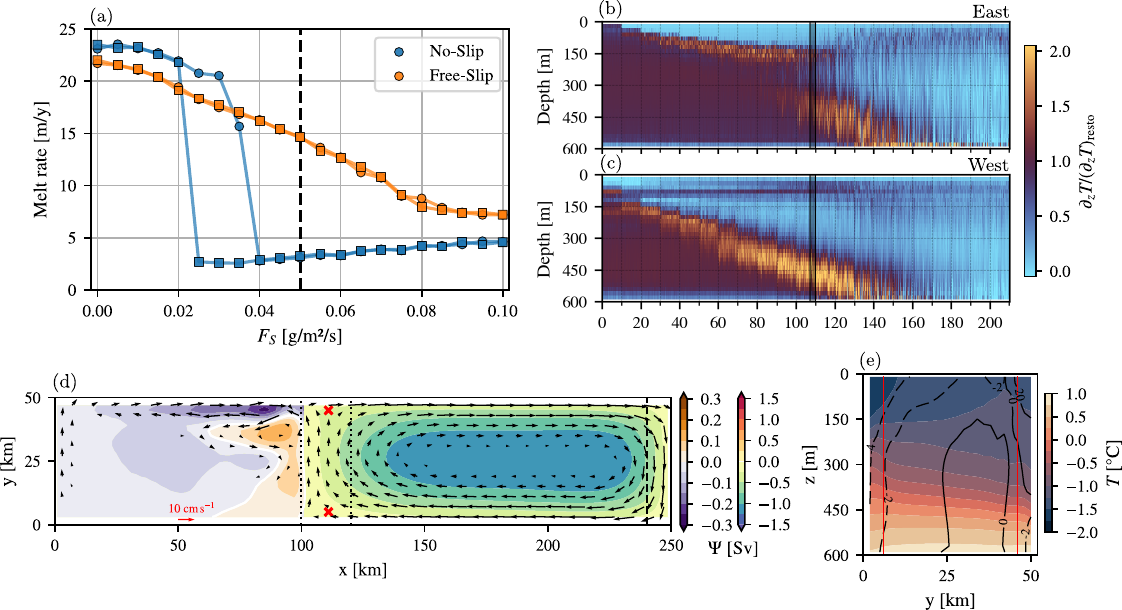}
\caption{Results of the quasi-static FSLIP experiment with free-slip lateral boundary conditions, $T_0=1\cels$, $\mathrm{SSS=34.0\psu}$, and $\mathrm{SBS}=34.5\psu$.
(a) Basal melt rates as a function of the surface salt flux (orange symbols). We also show the results for TEMP1 (blue symbols) for comparison. 
(b,c) Normalized monthly averaged temperature stratification plotted as a function of time, at $x=110$ km and $y=6$ km (eastern side) in (b), and $x=110$ km and $y=46$ km (western side) in (c). The thin shaded area in black in the middle represents the time window used for temporal averaging of the variables shown in (d) and (e). 
(d) Barotropic stream function. Red crosses show the positions of the temperature stratification profiles displayed in (b) and (c).
(e) $y-z$ slice of the temperature field at $x=110$ km. Black contours are meridional velocity contours [cm/s], negative (dashed) when directed towards the cavity, and positive (solid) when directed away from the cavity. Red verticals lines show the positions of the temperature stratification profiles displayed in (b) and (c). The black dashed line in (a) represents the surface salt flux imposed in (d) and (e).}
\label{FigFS}
\end{figure}

The ocean circulation at the front of the ice self is expected to play an important role on the stability of the water column beneath the polynya since it affects the transport of the meltwater coming from the ice-shelf cavity, the offshore water coming from the restoring region, and the DSW formed at the polynya surface.
We discuss the implications of modifying this circulation by changing the boundary conditions at the lateral walls of the spatial domain.

Figure~\ref{FigFS} compares the reference TEMP1 experiment with the FSLIP experiment (Table~\ref{table:exp}), wherein the lateral boundary conditions are changed from no-slip to free-slip.
The response of the ice shelf to surface salt forcing is completely different between the no-slip and free-slip cases. The basal melt rate transitions continuously and slowly from a warm mode to a cold mode, as shown in Fig.~\ref{FigFS}(a), and the transition is completely reversible as warm-to-cold and cold-to-warm experiments give the same response (cf. overlapping orange circles and squares).
The thermocline depth under the polynya shows a reversible evolution and is therefore only plotted for the warm-to-cold experiment, on the eastern side of the domain in Fig.~\ref{FigFS}(b), and on the western side in Fig.~\ref{FigFS}(c).
The mixed layer depth increases with time on both sides, as surface salt fluxes increases. However, the thermocline deepens first more slowly on the eastern side in comparison to the western side, and then relatively abruptly around year 100, beyond which it matches more closely the western side stratification. 
Therefore, it seems that the eastern and western sides of the polynya are not impacted by the same water masses fluxes.
Vertical dark bands are visible on both stratification plots in the layer above the thermocline at late times (cold conditions), meaning that the mixed layer is subject to episodic restratification events with warm water intrusions, which likely disrupt the cold mode dynamics.
To better understand the change of behavior between the no-slip and free-slip cases, the barotropic stream function of the free-slip experiment is shown in Fig.~\ref{FigFS}(d), for $F_S=0.05\,\mathrm{g/m^2/s}$, i.e. when the warm mode still prevails with free-slip conditions, but a cold mode circulation arises in the no-slip experiment (see Fig.~\ref{Fig3}).
Instead of counter rotating gyres along the $x$ axis, a single strong cyclonic gyre occupies the whole open ocean, from the front of the ice shelf to the restoring region. 
This modified circulation impacts the water properties at the front of the ice shelf in two ways.
Firstly, the meltwater releases on the western side are directly advected towards the restoring region where they transform into more saline water masses. Thus, the meltwater positive feedback is attenuated, and both circulation modes become less self-sustained.
Secondly, restoring temperature and salinity profiles are directly advected from the restoring region toward the polynya on the eastern side, favoring restratification, warm water intrusions and hence warm regime dynamics. The east-west asymmetry is further highlighted in Fig.~\ref{FigFS}(d), which shows a $y-z$ slice of the temperature field under the polynya. The direct connection between the restoring and polynya regions is responsible for the restratification events observed in Fig.~\ref{FigFS}(b,c), even when strong surface salt fluxes are applied. In the end, the free-slip experiments show reversible transitions because the domain-scale circulation suppresses the meltwater positive feedback.

\subsection{Impact of subgrid-scale dissipation: changing the horizontal viscosity}

\begin{figure}[ht!]
\centering
\noindent\includegraphics[width=1.\textwidth]{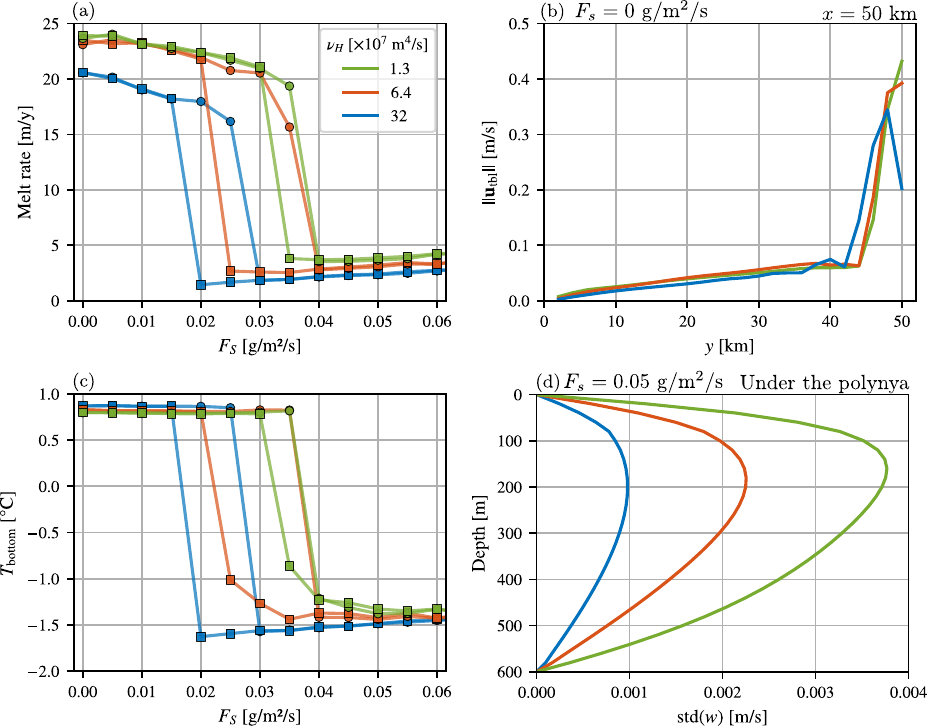}
\caption{Quasi-static experiments varying the horizontal viscosity $\nu_H$ and diffusivity $\kappa_H$ simultaneously ($\kappa_H=\nu_H$). The properties of the restoring region are $\mathrm{SSS}=34.0\psu$, $\mathrm{SBS=34.5\psu}$, $\mathrm{SST}=-1.9\cels$, and $T_0=1\cels$.
(a) Basal melt rates as a function of surface salt fluxes with either increasing $F_S$ (circles) or decreasing $F_S$ (squares).
(b) Norm of the top boundary layer horizontal velocity at $x=50$ km (300 m depth) for $F_S=0$ (warm mode).
(c) Ocean bottom temperature under the polynya as a function of $F_S$.
(d) Standard deviation of the vertical velocity below the polynya when $F_S=0.05\sfx$. 
}
\label{FigLDF}
\end{figure}

The choice of horizontal viscosity is not strongly constrained in ocean models, such that coefficients are often selected based on physical and numerical grounds that can vary substantially between model configurations~\cite{griffies2004FundamentalsOceanClimate}. Horizontal viscosity can yet directly influence tracer transport, e.g., in boundary currents~\cite{cessi1990DissipativeDynamicsWestern}, so in this section we investigate how variations in $\nu_H$ affect the transition thresholds and the stability of the two circulation regimes.

Figure~\ref{FigLDF} summarizes the viscosity-sensitivity experiments TEMP1, NUH1.3, and NUH32 listed in Table~\ref{table:exp}. For consistency, the horizontal diffusivity is modified simultaneously as $\kappa_H=\nu_H$.
We discuss the results of three quasi-static experiments having the same linear restoring profiles ($\mathrm{SSS}=34.0\psu$, $\mathrm{SBS=34.5\psu}$, $\mathrm{SST}=-1.9\cels$, and $T_0=1\cels$). The first experiment uses the same bilaplacian viscosity coefficient as in the original set of experiments (orange symbols and curve), the second one has a five times higher coefficient (blue), and the third one has a five times lower coefficient (green).
Note that we found it necessary to decrease the time step to 180 s for the two new experiments to ensure numerical stability~\cite{griffies2004FundamentalsOceanClimate}.

Fig.~\ref{FigLDF}(a) shows that the transition thresholds are affected by the viscosity coefficient, and that the bistable region becomes smaller when $\nu_H$ either increases or decreases.
In the most viscous simulation (blue symbols and lines), the basal melt rates are always weaker than in the other simulations. We demonstrate that the smaller basal melt rate and lower warm-to-cold threshold for the higher-viscosity simulation is due to a weakening of the meltwater plume. Fig.~\ref{FigLDF}(b) represents the norm of the horizontal top boundary layer velocity $\lVert \mathbf{u_\mathrm{tbl}}\rVert$, at $x=50$ km (mid-depth in the cavity), for $F_S=0$.
The meltwater boundary current is always visible close to the western boundary ($y=50$ km). However, its thickness and maximum strength vary with $\nu_H$. Specifically, it is thicker and weaker for the higher $\nu_H$ case. This result qualitatively agrees with the Munk layer scaling, which predicts a thickness proportional to $\left(\nu_H/\beta_\mathrm{topo}\right)^\frac{1}{5}$ and a transport in the layer proportional to $\left(\beta_\mathrm{topo}/\nu_H\right)^\frac{1}{5}$, where $\beta_\mathrm{topo}$ [$\mathrm{s^{-1}m^{-1}}$] is the topographic beta effect coefficient induced by the variable water column thickness in the cavity~\cite{munk1950WINDDRIVENOCEANCIRCULATION,cessi1990DissipativeDynamicsWestern,griffies2008FormulatingEquationsOcean}.
The increase of the horizontal viscosity thus lowers meltwater production through a decrease of $\gamma_T$, which weakens the meltwater feedback and ultimately lowers the warm-to-cold transition threshold, as shown in Fig.~\ref{FigLDF}(a).
We note that the zonal resolution $\Delta y = 2$ km of our configuration only partially resolves the western boundary layer. Thus, we would also expect the transition thresholds to depend on $\Delta y$. 

Fig.~\ref{FigLDF}(a) also shows that the cold-to-warm transition threshold decreases as $\nu_H$ increases. At lower horizontal viscosity, CDW penetrates more readily toward the cavity front due to stronger horizontal velocities (not shown). This is reflected in Fig.~\ref{FigLDF}(c), which shows that the bottom temperature beneath the polynya increases more rapidly prior to the cold-to-warm transition when $\nu_H$ is low (green symbols and line). The resulting increase in heat transport into the cavity enhances basal melting and meltwater production, thereby stabilizing the water column and favoring the warm mode. Consequently, the cold-to-warm transition occurs at larger values of $F_S$ as $\nu_H$ decreases. Figure~\ref{FigLDF}(d) further shows vertical profiles of the standard deviation of vertical velocity beneath the polynya for $F_S=0.05\sfx$, a forcing for which the cold mode dominates in all experiments. The vertical velocity fluctuations increase as $\nu_H$ decreases, indicating stronger vertical motions at low viscosity. However, this enhanced vertical activity does not stabilize the cold mode, since Fig.~\ref{FigLDF}(a) shows on the contrary that the cold regime becomes less stable as $\nu_H$ decreases. 
This suggests that the sensitivity of the cold-to-warm transition to $\nu_H$ is primarily controlled by changes in the meridional transport rather than by changes in vertical mixing.

\section{Discussion and conclusion}

In this study, we used the NEMO ocean model to investigate how salt rejection in coastal polynyas affects the stability of warm and cold ice-shelf cavity circulation regimes. We ran several hysteresis experiments to assess the reversibility of the transitions. Our configuration builds upon previous idealized numerical setups that did not include surface buoyancy forcing and thus focused primarily on the warm circulation regime~\cite{holland2008ResponseIceShelf,asay-davis2016ExperimentalDesignThree,yung2026ResultsSecondIce}. We highlight four key take-aways and discuss model limitations in the next paragraphs.

\textit{First take-away: we report irreversible transitions between cold and warm circulation regimes.} The emergence of a cold convective circulation regime for high enough surface salt fluxes aligns well with previous studies that demonstrated cold-to-warm regime shifts when the continental shelf freshen, based on both realistic~\cite{caillet2023DriversReversibilityAbrupt,hazel2020BistabilityFilchnerRonneIce,daae2020NecessaryConditionsWarm,song2025RegionalConditionsDetermine,jin2025CurrentReversalLeads,naughten2021TwotimescaleResponseLarge,mathiot2023SouthernOceanWarming,hellmer2017FateSouthernWeddell,haid2023DriversRegimeShifts,hoffman2024IceshelfFreshwaterTriggers,mathiot2026TransitionMuchWarmer} and low-dimensional models~\cite{hazel2020BistabilityFilchnerRonneIce,moorman2023CoastalPolynyasEnable,finucane2026TheoryMeltwaterPolynya,saddier2026IrreversibleTransitionsOcean}. The simplicity of our numerical configuration allowed to perform quasi-static experiments with surface salt fluxes increasing or decreasing incrementally, resulting in a succession of statistical steady states for each forcing value in the same fashion as ice-sheet modelling studies testing the reversibility of ice-sheet retreat~\cite{rosier2021TippingPointsEarly,garbe2020HysteresisAntarcticIce}.
We demonstrated that the system exhibits hysteresis with irreversible transitions for our reference configuration when the CDW temperature exceeds $0.5\cels$ and reversible behavior otherwise. The multistability is supported by a positive meltwater feedback that reinforces both regimes: high basal melt rates stabilize the water column in the warm mode, while low basal melt rates favor deep convection in the cold mode~\cite{hazel2020BistabilityFilchnerRonneIce,moorman2023CoastalPolynyasEnable,saddier2026IrreversibleTransitionsOcean}. Increasing the forcing temperatures amplifies this feedback, thereby widening the bistable range of surface salt fluxes.
We expect that increasing the exchange coefficients in the basal melt parameterization at the ice–ocean interface would similarly enhance this feedback and thus push the warm-to-cold transition threshold to larger surface salt fluxes.

\textit{Second take-away: the thermocline depth is an indicator of changes in melt and circulation regimes.} Warm-to-cold (resp. cold-to-warm) transitions occur exactly when the thermocline reaches (resp. lifts from) the seafloor.
Interestingly, early warning signals preceding regime transitions may be inferred from the thermocline dynamics: the variability of the thermocline depth increases (Fig.~\ref{Fig5}(a)) prior to warm-to-cold transitions and the frequency of intermittent restratification events at depth increases (Fig.~\ref{Fig5}(b)) before cold-to-warm regime shifts. In contrast, we did not find that basal melting rates carry useful information prior to transitions for early warning signals (Fig.~\ref{Fig4}(a,b)).
In the warm mode, the basal melt rate is modulated by the depth of the thermocline, with weaker basal melt rates when the thermocline deepens, in agreement with realistic studies of the Amundsen Sea~\cite{haigh2026WindDrivenCoastalPolynya, oconnor2025EnhancedWestAntarctic} and in a low-dimensional model of thermocline dynamics under the Amundsen Sea polynya~\cite{moorman2023CoastalPolynyasEnable}.
In the cold mode, the basal melt rate does not depend on the thermocline depth since convection occurs across the entire water column. Still, basal melt rates decrease with surface salt fluxes before the cold-to-warm transition, consistent with~\citeA{naughten2021TwotimescaleResponseLarge}, who link the basal melt rates to the barotropic circulation in the cavity.
Consistently, we find that surface salt fluxes create an anticyclonic gyre at the front of the ice shelf, which enhances basal melting in the upper part of the cavity~\cite{arthun2013EddyDrivenExchangeOpen}.
Interestingly, \citeA{jin2025CurrentReversalLeads} observe a current reversal at the front of the Amery ice shelf when transitioning from a cold to a warm mode, which we also have in our model with the reversal of the anticyclonic gyre. 
The quadratic melting law as a function of the restoring thermal driving predicted by \citeA{holland2008ResponseIceShelf} is recovered in the warm mode.
In the cold mode, the basal melt rate instead follows a linear scaling with thermal driving as DSW interferes with CDW, lowering the impact of far-field properties on ice losses. This result is in agreement with \citeA{burgard2022AssessmentBasalMelt} who found that quadratic melt models perform badly when they are forced with offshore water properties. In line with this, our linear scaling suggests that quadratic formulations may not always be appropriate for estimating basal melt rates from offshore conditions.

\textit{Third take-away: the buoyancy budget analysis provides an upper bound for the extent of the bistable region.} Rescaling the basal melt rates and the surface salt fluxes according to a buoyancy budget argument put forth in low-dimensional model studies~\cite{finucane2026TheoryMeltwaterPolynya,hazel2020BistabilityFilchnerRonneIce,moorman2023CoastalPolynyasEnable,saddier2026IrreversibleTransitionsOcean}, we found a reasonably good collapse of the hysteresis curves obtained with different restoring temperature forcing. However, the rescaling did not lead to the predicted single universal pair of transition thresholds. The bistable range for our simulations is  narrower than predicted as the salt flux thresholds for warm-to-cold transitions and salt flux thresholds for cold-to-warm transitions are lower and higher, respectively, than anticipated. The discrepancy can be attributed to the fact that the water column under the polynya is an open system, such that the buoyancy flux balance is not strictly between the meltwater outflow and air-ocean interface but also with the overall cavity and open ocean~\cite{xu2023PartI,xu2023PartII}. Similar observations have already been documented in numerical simulation studies that tried to predict transition thresholds from buoyancy budget arguments~\cite{hazel2020BistabilityFilchnerRonneIce,finucane2026TheoryMeltwaterPolynya}. We note that interconnections between multiple ice-shelf cavities exchanging meltwater fluxes could also impact the buoyancy budget and modify transition thresholds~\cite{hoffman2024IceshelfFreshwaterTriggers,silvano2018FresheningGlacialMeltwater}. 


\textit{Fourth take-away: far-field temperature, stratification and modeling choices impact the nature of the transitions.} We have explored several physical effects and modeling pathways that can change the reversibility of regime changes, which remains debated across realistic studies.
We have shown that the properties of the CDW intruding the continental shelf can impact the transition thresholds. 
On the one hand, low CDW temperature promotes lower warm-to-cold transition thresholds. On the other hand, high CDW density stratification promotes higher cold-to-warm transition thresholds. 
Weddell Sea ice shelves are subject to colder offshore CDW than the rest of Antarctica~\cite{schmidtko2014MultidecadalWarmingAntarctic,narayanan2023ZonalDistributionCircumpolar}
and Amundsen Sea offshore density profiles are more stratified than other regions of Antarctica~\cite{narayanan2023ZonalDistributionCircumpolar}.
The impact of temperature and stratification on reversibility acts in opposite direction for both the Amundsen Sea and Weddell Sea regions. Both transition thresholds should occur for relatively high $F_S$ values for the former and low $F_S$ values for the latter.
Therefore, it is difficult to conclude on a narrowing or a widening of the bistable zone due to the combined effect of far-field temperature and stratification, and on the reversibility of the transitions for these two Antarctic regions.
Other Antarctica regions, which are relatively warm and weakly stratified, can yet be expected to have a relatively large bistable region and thus more prone to irreversible transitions.
We have shown that the ocean circulation in front of the ice shelf can amplify the meltwater positive feedback by trapping the cavity outflow within a coastal gyre. In our idealized configuration, this mechanism promotes the persistence of both circulation regimes and ultimately contributes to the irreversibility of the transitions. The importance of local dynamics was demonstrated by comparing the no-slip simulations with analogue free-slip experiments (no friction on side walls), which do not feature a coastal gyre and experience continuous and reversible transitions.
\citeA{xu2023PartII} reached similar conclusions by showing that the presence of a headland or ice tongue adjacent to a polynya can induce a gyre that inhibits the export of DSW from the polynya region. More generally, several Antarctic ice shelves are associated with gyre-like circulations near their fronts. Examples include the Pine Island gyre in front of Pine Island ice shelf~\cite{poinelli2025SmallScaleHighFrequencyIce}, the Prydz Bay gyre adjacent to the Amery ice shelf~\cite{jin2025CurrentReversalLeads}, and the cyclonic and anticyclonic gyres over the Berkner bank and Filchner trough at the front of the Filchner ice shelf~\cite{naughten2019ModelingInfluenceWeddell}. These circulation features may increase the propensity of ice-shelf cavities to undergo irreversible transitions.
We finally investigated the role of the horizontal viscosity and showed that it can also modify the transition thresholds. Specifically, increasing the viscosity can reduce the heat flux from the continental shelf towards the cavity in the cold regime, modify the circulation in the ice-shelf cavity, and intensify vertical velocities associated with convective patterns. 
A future research avenue would be to test the impact of different choices for the horizontal and vertical subgrid scale dissipation schemes (related to both viscosity and diffusivity) in more realistic configurations.

\textit{Model limitations.} Our idealized configuration was well-suited for long quasi-static forcing experiments, which are typically inaccessible to realistic configurations. However, it lacks physical processes that are considered key to real cavities.
First, our configuration does not have a realistic topography for both the ice-shelf draft and the sea floor, even though topographical features, such as troughs, are expected to control circulation patterns and thus basal melting~\cite{derydt2024GeometricAmplificationSuppression,bradley2022InfluencePineIsland}.
It also omits continental shelf–break processes, which can modulate CDW inflows onto the shelf through (positive or negative) large-scale feedback~\cite{thompson2018AntarcticSlopeCurrent}. For example, \citeA{si2023HeatTransportAntarctic} showed that cross-slope salinity gradients across the shelf break impact the heat transport towards the continental shelf.
Another limitation stems from the idealized forcing scheme imposed at the sea surface, which omits surface wind stresses and discards sea-ice coupling. 
The modeling of surface conditions may yet have a leading order effect on dynamics as \citeA{xu2023PartI} showed that the wind driven circulation impact the formation and advection of DSW at the front of the ice shelf, and that the extent of the polynya is strongly dependent on the interplay between the sea ice and wind stresses.
Freshwater fluxes coming from other ice-shelf cavities could also impact the buoyancy budget at the front of the ice shelf~\cite{hoffman2024IceshelfFreshwaterTriggers,silvano2018FresheningGlacialMeltwater}.
Finally, our study focuses on the steady response of the ice shelf to coastal polynya forcing, which is an appropriate framework to characterize feedback processes and the nature of regime shifts, but leaves open the question of time scales. 
Nevertheless, we remark that our 10-year steps were sufficient to capture full regime transitions, giving a first order-of-magnitude estimate for the transition time scales in our specific configuration.
We expect that the flushing time scales of the cavity~\cite{holland2017TransientResponseIce}, and the deepening and lifting time scales of the thermocline~\cite{arthun2013EddyDrivenExchangeOpen,sohail2020DynamicsMixedLayer,xu2023PartI} are key quantities to consider when exploring the abruptness of regime shifts.

\section*{Open Research Section}
Data and software used in this study are openly available. The analysis scripts and code supporting the results
of this paper, as well as the processed dataset generated during the current study, are archived in Zenodo~\cite{saddier2026data}.

\section*{Conflict of Interest disclosure}
The authors declare there are no conflicts of interest for this manuscript.

\acknowledgments
The numerical simulations were carried out using the resources of the \textit{Pole Scientifique de Mod\'elisation Num\'erique \& Centre Blaise Pascal} at ENS de Lyon, and in particular the SIDUS system~\cite{quemener2013sidus}. We are grateful to Cerasela Calugaru for her help with the platform.
This work is funded by the European Union. Views and opinions expressed are however those of the authors only and do not necessarily reflect those of the European Union or the European Research Council Executive Agency. Neither the European Union nor the granting authority can be held responsible for them. This work is supported by an
ERC grant (IceAblation, 101117317).
Christopher Bull and Louis Saddier acknowledge support from the UK Research and Innovation (UKRI) through the Rethinking Antarctic Sea Level Projections (RASP) project (NE/Y001451/1).
This work is also supported by the Fondation Simone et Cino Del Duca—Institut de France.

%
%

\bibliography{references}

%
%
%
%
%

\appendix

\section{Model parameters and key variables}

\begin{table}
\label{tab:parameter}
\caption{List of model parameters and key variables. We provide either a standalone value or a range of values (in between brackets) for parameters. We do not provide values for variables, which are model output, since they are not prescribed.}
\centering
\begin{tabular}{c|l c r l }
  & \textbf{Parameter}  & \textbf{Symbol} & \textbf{Value} & \textbf{Unit}  \\
 \hline
 \textbf{Ice-ocean} & Freshwater flux  & $q_\mathrm{fw}$ & - & $\mathrm{kg\,m^{-2}\,s^{-1}}$ \\
 \textbf{forcing} & Liquidus slope  & $\lambda_1$ & $-0.0573$ & $\mathrm{^\circ C\,(g\,kg^{-1})^{-1}}$ \\
 & Liquidus intercept  & $\lambda_2$ & $0.0832$ & $\mathrm{^\circ C}$ \\
 & Liquidus depth coefficient  & $\lambda_3$ & $7.59\times10^{-4}$ & $\mathrm{^\circ C\,m^{-1}}$  \\
 & Ice-ocean interface temperature & $T_b$ & - & $\mathrm{^\circ C}$ \\
 & Ice-ocean interface salinity  & $S_b$ & - & $\psu$  \\
 & Ice-ocean interface depth & $H_\mathrm{isf}$ & $[90,555]$ & $\mathrm{m}$ \\
 & Heat capacity of seawater & $c_p$ & $3974$ & $\mathrm{J \, ^\circ C^{-1} \, kg^{-1}}$ \\
 & Latent heat of fusion & $\mathcal{L}$ & $3.34\times10^5$ & $\mathrm{J \, kg^{-1}}$ \\
 & Losch layer thickness  & $h_\mathrm{tbl}$ & $30$ & $\mathrm{m}$ \\
 & Losch layer temperature & $T_\mathrm{tbl}$ & - & $\mathrm{^\circ C}$ \\
 & Losch layer salinity & $S_\mathrm{tbl}$ & - & $\mathrm{g\,kg^{-1}}$ \\
 & Losch layer density  & $\rho_\mathrm{tbl}$ & - & $\mathrm{kg\,m^{-3}}$ \\
 & Losch layer horizontal velocity  & $\mathbf{u}_\mathrm{tbl}$ & - & $\mathrm{m\,s^{-1}}$ \\
 & Tidal velocity  & $U_\mathrm{tide}$ & $0.01$ & $\mathrm{m \, s^{-1}}$ \\
 & Friction velocity & $u_*$ & - & $\mathrm{m\,s^{-1}}$ \\
 & Turbulent heat transfer velocity  & $\gamma_T$ & - & $\mathrm{m\,s^{-1}}$ \\
 & Turbulent salt transfer velocity  & $\gamma_S$ & - & $\mathrm{m\,s^{-1}}$ \\
 & Turbulent heat transfer coefficient  & $\Gamma_T$ & $2.15\times10^{-2}$ & - \\
 & Turbulent salt transfer coefficient  & $\Gamma_S$ & $6.14\times10^{-4}$ & - \\
\hline
\textbf{Surface} & Polynya salt surface flux & $F_S$ & $[0,0.075]$ & $\mathrm{g\,m^{-2}\,s^{-1}}$ \\
\textbf{forcing} & Sea surface temperature  & $\mathrm{SST}$ & - & $\mathrm{^\circ C}$ \\
 & Surface freezing temperature & $T_s$ & $-1.9$ & $\mathrm{^\circ C}$ \\
 & Surface heat restoring coefficient & $K$ & $-40$ & $\mathrm{W\,m^{-2}\,^\circ C^{-1}}$ \\
\hline
\textbf{Restoring} & Bottom restoring temperature & $T_0$ & $[-1.9,2]$ & $\mathrm{^\circ C}$ \\
\textbf{forcing} & Restoring temperature profile & $T_\mathrm{resto}$ & $[-1.9,T_0]$ & $\mathrm{^\circ C}$ \\
 & Restoring salinity profile & $S_\mathrm{resto}$ & $[34,34.5]$ & $\mathrm{g\,kg^{-1}}$ \\
 & Restoring frequency & $f_\mathrm{resto}$ & $[0,0.1]$ & $\mathrm{day^{-1}}$ \\
\hline
\textbf{Ocean} & Thermal expansion coefficient & $\alpha$ & $3.733\times10^{-5}$ & $\mathrm{^\circ C^{-1}}$ \\
\textbf{dynamics} & Salinity contraction coefficient & $\beta$ & $7.843\times10^{-4}$ & $\mathrm{(g\,kg^{-1})^{-1}}$ \\
 & Top/bottom drag coefficient & $C_D$ & $2.5\times10^{-3}$ & - \\
 & Horizontal viscosity & $\nu_H$ & $6.4\times10^7$ & $\mathrm{m^4\,s^{-1}}$ \\
 & Horizontal diffusivity & $\kappa_H$ & $6.4\times10^7$ & $\mathrm{m^4\,s^{-1}}$ \\
 & Stable vertical viscosity & $\nu_v$ & $10^{-3}$ & $\mathrm{m^2\,s^{-1}}$ \\
 & Stable vertical diffusivity & $\kappa_v$ & $5\times10^{-5}$ & $\mathrm{m^2\,s^{-1}}$ \\
 & Convective vertical viscosity & $\nu_v^c$ & $0.1$ & $\mathrm{m^2\,s^{-1}}$ \\
 & Convective vertical diffusivity & $\kappa_v^c$ & $0.1$ & $\mathrm{m^2\,s^{-1}}$ \\
 & Coriolis frequency & $f$ & $-1.4\times10^{-4}$ & $\mathrm{s^{-1}}$ \\
 & Horizontal resolution & $\Delta x=\Delta y$ & $2$ & $\mathrm{km}$ \\
 & Vertical resolution & $\Delta z$ & $20$ & $\mathrm{m}$ \\
 & Time resolution & $\Delta t$ & $360$ & $\mathrm{s}$ 
\end{tabular}
\end{table}

\end{document}